\documentclass[10pt,twocolumn,times]{article}
\usepackage{latex8}
\usepackage{graphicx}

\newcommand{\ie}{\emph{i.e.}}
\newcommand{\eg}{\emph{e.g.}}

\newcommand{\ndocs}{N_{\mathit{D}}}

\newcommand{\ndimensions}{n_{\mathit{Dimensions}}}

\newcommand{\fixme}[1]{}
\newcommand{\comment}[1]{}
\newcommand{\doit}[1]{}
\newcommand{\henrich}[1]{}
\newcommand{\stichpunkte}[1]{}
\newcommand{\mathvector}[1]{\mathbf{#1}}

\newcommand{\ideen}[1]{}

\title{Comparison of Image Similarity Queries in P2P Systems}
\author{Wolfgang M\"uller\\
Media Informatics\\
Bamberg University, Germany\\
wolfgang.mueller@wiai.uni-bamberg.de\\
\and
P. Oscar Boykin\\
Dept. of Electrical and Computer Engineering\\
University of Florida\\
Gainesville, Florida 32611-6200\\
boykin@ece.ufl.edu\\
\and
 Nima Sarshar\\
Electrical Engineering Dept.\\
University of California\\
Los Angeles, California 90095\\
        nima@ee.ucla.edu\\
\and
Vwani P. Roychowdhury\\
Electrical Engineering Dept.\\
University of California\\
Los Angeles, California 90095\\
vwani@ee.ucla.edu}

\begin{document}

\maketitle

\begin{abstract}

Given some of the recent advances in Distributed Hash Table (DHT) based
Peer-To-Peer (P2P) systems
we ask the
following questions: Are there applications where unstructured queries are
still necessary (i.e., the underlying queries do not efficiently map onto any
structured framework), and are there  unstructured P2P systems that can deliver
the high bandwidth and computing performance necessary to support such
applications.  Toward this end, we consider an image search application which
supports queries based on image similarity metrics, such as color histogram
intersection, and discuss why in this setting, standard DHT approaches are not
directly applicable.  We then study the feasibility of implementing such an
image search system on two different unstructured P2P systems: power-law
topology with percolation search, and an optimized super-node topology using
structured broadcasts. We examine the average and maximum values for node
bandwidth, storage and processing requirements in the percolation and super-node models,
and show that current high-end computers and high-speed links have sufficient
resources to enable deployments of large-scale complex image search systems.
\end{abstract}

\section{Introduction}

The first widely known pure P2P system that tried to bring Napster-like
functionality was the unstructured P2P system Gnutella. The overlay
network created by Gnutella's peers forms a random graph, where search was
mostly done via complete flooding of the network.  Its imperfections have
spawned extensive research. Much of this research is directed towards DHTs
(\emph{e.g.}~\cite{MaM2002,SMN2001}), and distributed indexing structures
derived from DHTs (\emph{e.g.} \cite{TXM2002}), super
peer architectures (\emph{e.g.} \cite{YaG2003}) and approaches
improving the link structure of P2P networks (\emph{e.g.}~\cite{NgS2003}).

DHTs excel at key-value lookup because that is the basic primitive
of the hash table data structure.  However, hash tables are not the
most efficient data structure for all algorithms. In this work we
are interested in storing images, which may be thought of as vectors
of large dimension, and searching to find images which are close to
a query image by some given metric.  This work compares content
based image retrieval (CBIR) in structured against unstructured P2P
systems.  Unstructured systems can answer completely general
queries, since there is no structure imposed on the data. We ask if
sacrificing the flexibility of unstructured systems for structured
P2P systems results in a reduction of query bandwidth for the case
of CBIR.

Within this article, we treat \emph{content-based image retrieval as
  an application scenario}. We think that this scenario is interesting
for P2P in several respects. (1) \emph{We believe that there is a need
  for such applications.}  The recent success of image blogs and image
sharing servers like \texttt{flickr.com} has shown that people have
the wish to share and to publish their images. (2) \emph{Current
  widespread methods of indexing such images are unsatisfactory.} In
\texttt{flickr.com}, the images are searchable by annotation.
Unfortunately the annotation quality is low (as will be described
more in-depth below). Organizing images by time~\cite{GGP2002} is
useful when indexing one's own collection where  images can usually
be grouped into images pertaining to events that are interesting to
the user, but it becomes barely useful when considering collections
that grow each second by at least one image.  (3) \emph{Alternative
methods
  are expensive.}  Sophisticated content-based image indexing methods
that extract visual features from the items to be queried need a large
amount of processing power, and processing queries is
expensive\footnote{That means, they typically require many disk
  accesses, much processing power and as a consequence much time to be
  processed.}  compared to processing text queries.

While CBIR is deemed unsatisfactory as a complete indexing solution
for images, \emph{the conviction underlying this paper} is that CBIR
methods are going to be useful for improving \texttt{flickr.com}
like systems. Bad image annotation is not going to go away. So, we
seek systems that help the user in the case that good annotation is
not present. We are also convinced that people will be interested in
\emph{combined rankings}, which combine visual aspects, surrounding
text, etc. into one common measure. Unless one wants to restrict
oneself to systems that first evaluate similarity with respect to
text and simple meta-data before refining the search using CBIR and
other complex methods, one will need to come up with good methods
that are able to process CBIR queries.

This paper presents two findings.
First, we find
that unstructured systems are efficient enough in terms of communication
and computational costs for the CBIR application
we consider to scale to millions of users.
Second, in the case of image similarity search, current structured systems
offer no advantage over unstructured systems.
This paper is organized as follows.  Section \ref{sec:nn_hd} discusses prior
work on the nearest-neighbor search problem in high dimensional spaces, a
general version of the problem we consider in this work.  Section
\ref{sec:setting} describes the particular case we are focusing on:
image similarity search at the scale of approximately one 
million users.  Section \ref{sec:supernode} studies the cost of implementing
an image search system using a supernode architecture and Section
\ref{sec:perc} studies the cost for a percolation search based architecture.
Finally, in Section \ref{sec:compare_utos} we compare the unstructured
image search systems to a structured image search system and find
that structured search offers no benefit for this case.

\section{Nearest neighbor search in high dimension}
\label{sec:nn_hd}

The nearest-neighbor search problem is the following: given a set
of points in a metric space $P$, for a given query point $Q$,
find an element $x$ of $P$ such that $d(x,Q)\le d(y,Q)$ for all
$y\in P$.

The classic way of performing CBIR is to extract real-valued feature
vectors from images and then map the search problem to the problem of
finding the $k$ nearest neighbors ($k$-NN) to the query vector. For
$d<\approx 10$ there exist centralized data structures that find the
$k$-NN in $O(\log N)$ time for collections of size $N$.  However,
literature on non-distributed indexing has observed~\cite{WSB1998} that
\emph{exact} search in high-dimensional data is very hard, due to the
so-called \emph{curse of dimensionality}. Due to the curse of dimensionality
tree-based indexing structures break down in the sense that in
realistic scenarios $O(N)$ nodes need to be visited before finding the
\emph{exact} $k$-NN. For non-distributed ---disk based--- indexing
structures one interesting and well-known solution~\cite{WSB1998}
consists in rendering full scan queries more efficient using a full
table scan approach. This algorithm  exhibits $O(N)$ complexity just as a tree-based indexing structure in high dimensions,
however, the absolute query duration is reduced with respect to the
tree-based solution.

Though not the subject of this work, one may also consider the
approximate version of the nearest-neighbor search problem. The
approximate nearest-neighbor search problem is the following: given
a set of points in a metric space $P$, for a given query point $Q$,
and a slackness parameter $\epsilon$ find an element $x$ of $P$ such
that $d(x,Q)\le (1+\epsilon)d(y,Q)$ for all $y\in P$.  An efficient
algorithm for the approximate nearest-neighbor search problem is
known for several common metric spaces \cite{KOR2000}. 
It is an interesting open problem to see if it
can be efficiently adapted to
a distributed system.

Other proposals include using geometric dimensionality reduction techniques.
Kleis and Zhou provide a review of relevant results in the context of P2P
networks in \cite{KZ2004}.

\subsection{P2P approaches to nearest-neighbor search}
Clearly, finding the set $\{x | d(x,Q) \le \delta\}$ for a given
$Q,\delta$ is an embarrassingly parallel problem.  If one spreads
the database over a P2P network, one gets the full benefit of parallelization.
In the absence of an efficient exact algorithm for the nearest-neighbor
problem in high dimension, this may be the best one can do.  Indeed
we consider this approach in Sections \ref{sec:supernode} and \ref{sec:perc}.

In addition to the above, one can also attempt to use some structured
P2P network to reduce the number of nodes that must be contacted to
execute a query.  We describe one such approach, PRISM below. The
literature on P2P indexing using structured networks is very broad. As
one starting point for reading we suggest~\cite{TXM2002}.

\subsubsection{PRISM}

PRISM indexes each vector $\mathvector{x}$ by placing $\mathvector{x}$
on a small number of nodes in a Chord DHT. The placement of the vector
is calculated using distances to a fixed set of \emph{reference
  vectors}.  When processing a query, the node issuing the query
$\mathvector{q}$ calculates the set of nodes where $\mathvector{q}$
would be placed and searches for similar nodes there, sending them
$\mathvector{q}$ as the query. The main innovation of PRISM is the
algorithm for finding the nodes on which to place the data vectors.

In order to index a vector $\mathvector{x}$, the distance of
$\mathvector{x}$ to a number $n_r$ of reference vectors
$\mathvector{r_i}$ is calculated, yielding
$\mathbf{\delta}:=(\delta_1\delta_1,\ldots,\delta_{n_r}):=(\delta(\mathvector{x},\mathvector{r_1}),\ldots,\delta(\mathvector{x},\mathvector{r_{n_r}}))$.
Then the $\mathvector{r_i}$ are \emph{ranked} by their similarity.
The result of this ranking is a list of indices
$\mathvector{\iota}=(\iota_1,\ldots,\iota_{n_r})$ such that
$\mathvector{r_{\iota_1}}$ is the reference vector closest to
$\mathvector x$, $\mathvector{r_{\iota_2}}$ the second closest and
so on.

Then, pairs of indices are formed. The pair formation is a fitting
parameter, the original PRISM paper suggests $\{\iota_1,\iota_1\},
\{\iota_1,\iota_2\}, \{\iota_2,\iota_3\}, \{\iota_1,\iota_3\},
\{\iota_1,\iota_4\},\linebreak \{\iota_2,\iota_5\}, \{\iota_2,\iota_4\},
\{\iota_3,\iota_4\}, \{\iota_1,\iota_5\}, \{\iota_4,\iota_5\},
\{\iota_3,\iota_5\}$ for their dataset. From each of the pairs a Chord
key is calculated, and this key is used for inserting the vector
$\mathvector{x}$ into the Chord ring.

As was hinted above, query processing works by finding out which
peers would receive the query vector if it was a new data item and
forwarding the query vector to these peers. This involves, again,
the calculation of index pairs, which we will call \emph{query
pairs} in the following. In order to reduce query processing cost,
the query processor can choose to contact only nodes pertaining to
only a subset of the query pairs. Doing this also reduces recall, so
there is a tradeoff.

\section{An image search system}
\label{sec:setting}

Within this section, we describe Flickr, a popular web-based photo
sharing application. This application is currently immensely
popular. At the same time, one could easily imagine extending its
functionality towards content-based search.  By examining Flickr we
estimate the load for a P2P photo sharing system which we call
Plickr.

\subsection{About Flickr}
\texttt{flickr.com} gives members the opportunity to share photos among the public,
friends and family. Flickr members are allowed to comment on photos
they can see and to annotate them in a collaborative fashion.
Recently, \texttt{flickr.com} has experienced explosive growth of
popularity. As of the time of writing, \texttt{flickr.com} contains
about 40 million images, most of them publicly accessible. We
estimate that about 2 million users share images via
\texttt{flickr.com}.

One of the reasons for our interest into \texttt{flickr.com} is that
it has a SOAP-like API that allows easy access. 
This simplifies enormously building third party tools, as well as crawlers. As the user
structure is quite similar to what many people would like to have in
P2P file sharing systems (people peacefully sharing data they actually
own) we simply extrapolate from user behavior on \texttt{flickr.com}
to the behavior they would have in a P2P network.

Using \texttt{flickr.com}, we do not
obtain data about the online times of the users. However,
information about who shares
how much is already useful. 



\begin{figure}[t]
  \centering
  \includegraphics[width=2.7cm]{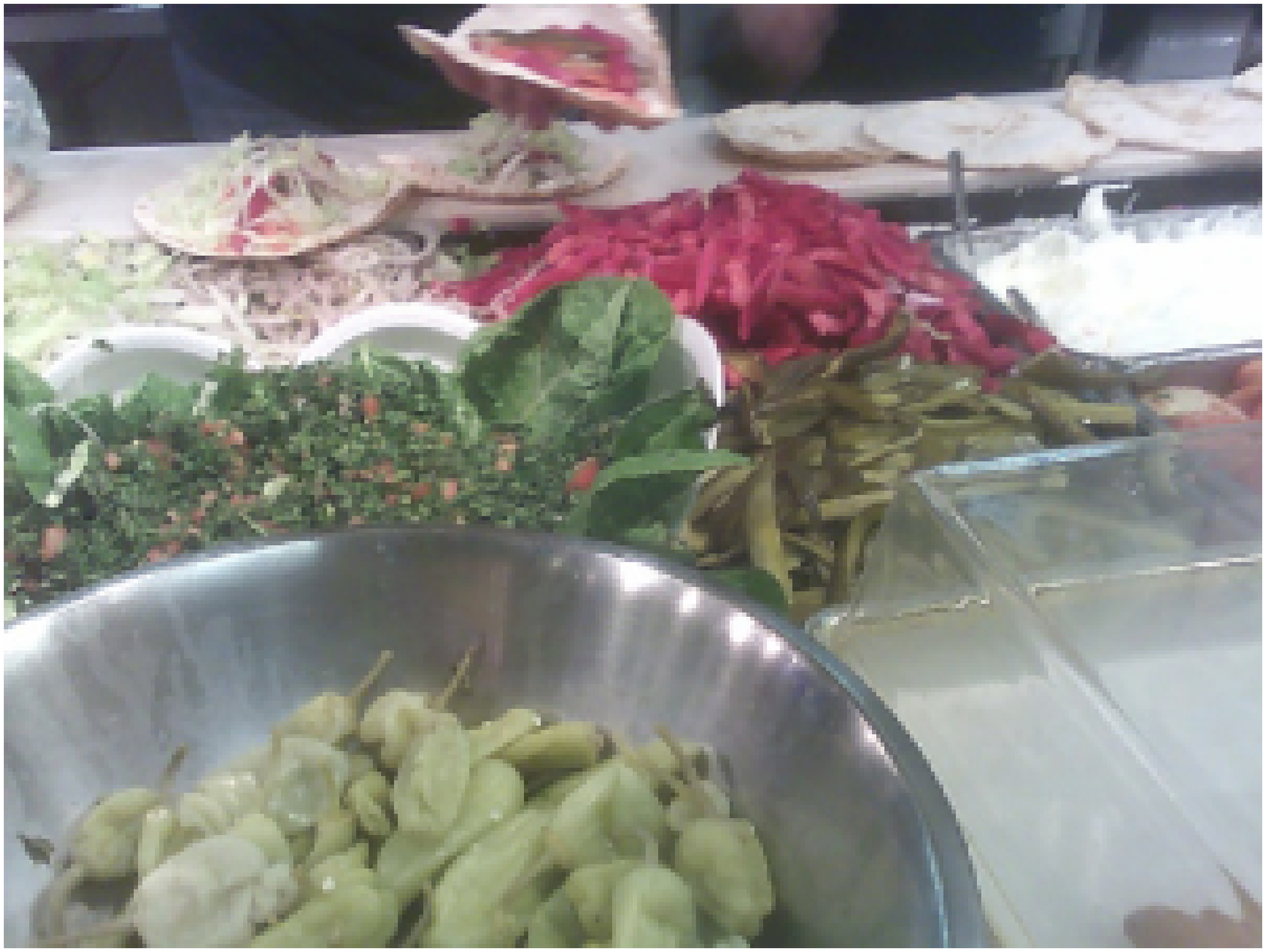}
  \includegraphics[width=2.7cm]{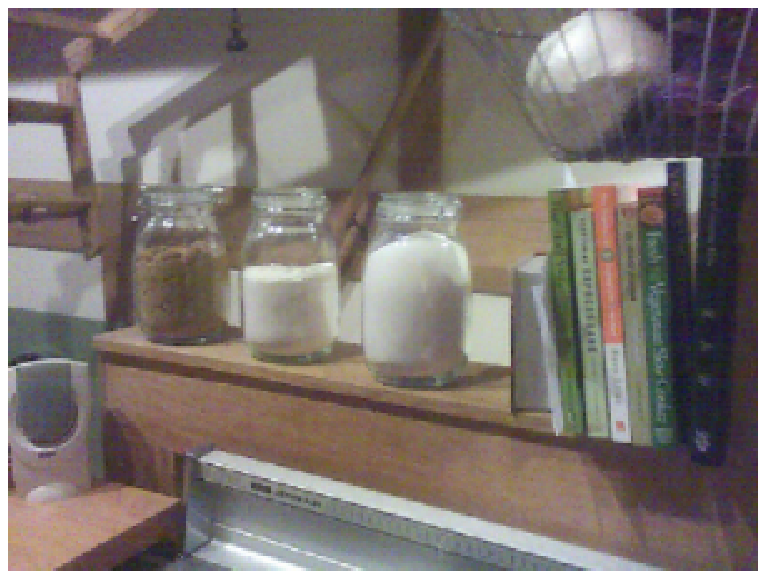}
  \caption{Example images tagged with the annotation tag
    \emph{phone}, taken from \texttt{flickr.com} with permission from Katy Wortman.}
  \label{fig:examples}
\end{figure}

One of the most interesting features of Flickr's is that members can
annotate other people's images. This leads to the surprising fact that
most of Flickr's images are annotated. However, as it is made simple
to add annotation to images by default (on a user-by-user basis), the
quality of the annotation is varying. For example, many Flickr members
have a large fraction of their photos annotated with the tag
\texttt{phone}.  This tag describes how the image came to Flickr, by a
camera built into a cellular phone. However, it is only rarely an
accurate description of the images' content, as the examples in
Fig.~\ref{fig:examples} show.
While the tag \emph{phone} is clearly the most extreme case of
annotation that carries little valuable information, it shows that
just citing the \emph{number} of images that are annotated does not
permit assessing the usefulness of this annotation.

\subsection{Plickr: content-based Flickr over P2P as a scenario for P2P-CBIR evaluation}

The above \emph{ad-hoc} assessment of annotation quality motivates the
view that it would be interesting to combine the search by annotation
tag (as offered by Flickr) by search based on image similarity as
provided by Content Based Image Retrieval systems (CBIRS). While CBIRS
 are unable to do true object recognition, they capture visual
similarity by translating each image into a data representation that
captures mainly image statistics and in some cases the spatial
relation of ``interesting'' regions. While using such features as the
sole image retrieval method is currently deemed unsatisfactory,
CBIRS are still the method of choice when no annotation is present.

Content Based Image Retrieval is computationally costly. Features need
to be extracted, and similarity search is much harder than similarity
search on text because the number of features
taken into account for obtaining a retrieval result is usually much
higher than the number of keywords in a keyword query.  
For more information about CBIR we point to an
often-cited overview paper \cite{SWS2000}.

P2P becomes particularly interesting through the fact that P2P-CBIR
potentially will make use of both the huge storage capacity and the
huge computing capacity distributed in the network. 

Let us consider a P2P network, in which each peer owner shares his or
her own photos with other users of the same P2P network. The network
would offer Flickr's functionality plus CBIR query by example
functionality. However, in contrast to Flickr, its inner workings
would be based entirely on P2P principles. We
will call this hypothetical network the \emph{plickr}
network
within this paper.

\subsection{Deriving a load scenario for \emph{plickr}}
\label{sec:plicker_load}

We assume the images contained in one Flickr user account to be a good
model for the images shared by one plickr peer.  Let us assume 1,000,000
users and thus 1,000,000 peers in our plickr
network. Our Flickr crawls ($\approx$2,000,000 images)%
indicate that the average user who
shares at least one image publicly shares on average 20 images.

The same as Flickr, plickr members query the data collection for
interesting images from time to time. Furthermore, we assume each
user performs 10 queries on average per day. We feel that is
reasonable as querying image is an exploratory process, so each
querier is likely to perform a query process consisting of multiple
queries. So, even if such a query process is performed less than
once per day by each user, we are likely to reach the said average
load.

\label{sec:features}
In this paper we will concentrate on a very simple way of performing
CBIR: retrieval by color histograms. They are known to provide a good
retrieval performance (\ie~result quality) for comparatively little
computing power and are the pet feature extraction method for indexing
structure evaluations.

Color histograms are obtained by cutting the color space in regions.
The color histogram then is a vector that contains one value
corresponding to each color space region. Each value of the histogram
expresses for the corresponding color region the probability that a
pixel drawn from the image falls into the color region.  This probability is estimated by simply counting pixels falling in each color region. The
usefulness of color histograms for CBIR depends on the color space
chosen and the way it is split into regions. For our load assumption,
we assume John R.~Smith's 166-D HSV histograms described in \cite{Smi1997}.
A wasteful but simple
representation would be of type \verb,float[166],. 


The classic way of evaluating the similarity of two histograms is the
\emph{histogram intersection}, however, the testing ground for most
CBIR indexing algorithm is the Euclidean distance which is why we
focus that distance measure.

\section{Search with unstructured P2P}
\label{sec:unstructured}

Unstructured P2P systems have one major advantage over
structured systems: once designed, implemented and deployed an unstructured
system can generally be used for any kind of query just by plugging in new
query processing.  There is no need to define new routing algorithms, 
network topologies, or caching strategies every time a new type of data or
query is introduced to the network.  On the other hand, structured P2P
systems \emph{may} reduce search complexity only if the search algorithm can be
efficiently mapped onto the topology of the structured network.

In this section, we compute the costs in bandwidth, computational resources,
and storage to use unstructured P2P for two models: a super-node system, and a
percolation search system.
After computing the costs of the system, in order to estimate
feasibility, we make some assumptions about the usage of the image
search system described in Section \ref{sec:setting}. We assumed
that each content item is a \texttt{float[166]} array, which uses
$166\times 4 = 664$ bytes of space. As we mentioned in  Section
\ref{sec:plicker_load}, every Flickr user inserts on average $C=20$
items into the network. Calculating the distance takes $f=332$ floating point
operations. Finally, we will assume that there are $N=2^{19}\approx
500,000$ users, and that each user will make $10$ queries per day or
$R=\frac{10}{24\times 60\times 60}=1.2\times 10^{-4}$ queries per
second. We assume each query and content requires $z=800$ bytes
(enough to hold the float vector and some routing information or
image meta-data).

\subsection{Super-node P2P networks}
\label{sec:supernode}

In the super-node model there are two types of nodes: leaf nodes and
super-nodes.  Each leaf node connects to a super-node and caches all its content
on that super-node (the super-node does not need to cache its own content). 
The leaf nodes require very
little resources, but super-nodes incur the maximum penalty.  The only
parameter in the system is $s$, the fraction of nodes which are super-nodes.

\begin{table}[t]
  \centering
  \caption{Nomenclature used throughout the paper}
{\small
  \begin{tabular}[t]{|c|p{6cm}|}
    \hline
    Symbol & Meaning \\\hline\hline
    $N$ & Total number of peers \\\hline
    $\left<k\right>$ & Expected degree of nodes within network\\\hline
    $p_m$ & relative frequency of nodes with degree $m$ \\\hline\hline
    $s$ & Fraction of super peers \\\hline
    $R$ & Query rate issued per peer ($1.2\times 10^{-4}\ s^{-1}$)\\\hline
    $C$ & Number of content items \emph{contributed} per peer ($20$) \\\hline
    $f$ & Number of operations to compare two vectors ($332$ FlOp)\\\hline
    $z$ & Size of each vector ($800$ B)\\\hline
    $B_{max,ave}$ & max/avg bandwidth required per peer\\\hline
    $P_{max,ave}$ & max/avg processing required per peer\\\hline
    $D_{max,ave}$ & max/avg disk space required per peer\\\hline
  \end{tabular}
  \label{tab:nomenclature}
}
\end{table}

\subsubsection{Resource requirements}
To compute average bandwidth, we count total copies of the queries
and divide by the total number of nodes.  We should note that this
metric is not very meaningful since no nodes see the average.  Leaf
nodes see almost no traffic, while super-nodes see the maximum
traffic.

Since each query is copied to $sN$ super-nodes plus
the leaf node that initiated the query, the average bandwidth is clearly,
$B_{ave} = \frac{RN (sN+1)z}{N} = RNz (s+ \frac{1}{N})\approx RzsN$.
All the super-nodes see the same bandwidth since all queries pass through them.
If we assume that the query crosses each edge in the multicast tree
(using an approach similar to \cite{NWS2003})
then it is necessary and sufficient for the maximum degree to be $3$,
thus $B_{max} = 3RzN $, which is independent of $s$.

Since there are $CN$ total content items, the average disk space is
$D_{ave} = CNz/N=Cz$.  Since all content is stored on the supernodes,
the maximum disk space is $D_{max} = CNz/Ns = Cz/s$

In the super-node system, each content is only copied (at most) one time.
The average processing requirement
is not very meaningful since like average bandwidth, no node experiences
this load.  Nodes either see almost no load, or maximum load.
We assume a linear complexity for search, so that $P= R D(f/z) N$:
$P_{ave} = R D_{ave}(f/z) N = RCfN$ .

Since, all the queries are processed by the super-nodes, we only need to
compute the number of content items on each super-node, and then multiply
by the query rate: $P_{max} = R D_{max}(f/z) N = \frac{CN}{sN}fRN = \frac{RCfN}{s}$.

\subsubsection{Trade-offs and numerical values}
\label{sec:sn_num}

One interesting feature of this model is that $P_{max}B_{ave} = f z C R^2 N^2$,
which is independent of $s$.  So, there is a trade-off between average
bandwidth and maximum CPU utilization. 
In the interest of considering some numerical values, we will
set $s=1/\sqrt{N}$, which means each super-node has as many leaf nodes as there
are super-nodes.  In practice, one will probably prefer to minimize bandwidth
to the extent that it is possible for the super-nodes to handle the load.

In the following table, we present performance metrics for the
super-node algorithm with $s=1/\sqrt{N}$, $N=2^{19}$ and the values of
$R,C,f,z$ from Table \ref{tab:nomenclature}.

\begin{tabular}{lll}
$B_{ave}=$ & $Rz\sqrt{N}$ &$ = 70 B/s = 560 bps$\\
$B_{max}=$ & $3RzN$ & $= 150,000 B/s = 1.2 Mbps$\\
$D_{ave}=$ & $Cz$ & $ = 16 KB$\\
$D_{max}=$& $Cz\sqrt{N}$ & $= 11 MB$\\
$P_{ave}=$ & $RCf N $ &$= 420 kFlOp/s$\\
$P_{max}=$ & $RCf N\sqrt{N}$ & $= 300M FlOp/s$
\end{tabular}

The values look reasonable.
Since modern CPUs have processing
power on the order of $4$ GFLOPS, the above processing requirements
are not more than one CPU.
The figure we might be most concerned about is $B_{max}$, however
that value is independent of $s$.
Now we compare the above with the percolation search
algorithm for unstructured networks.

\subsection{Percolation search in power-law networks}
\label{sec:perc}
In \cite{SBR2004} and subsequent work, the
authors show that using a combined random walk data
replication/random walk query distribution scheme (to be detailed
below) one can achieve sub-linear (in $N$) query complexity in
power-law networks.  Below we summarize this algorithm.

The degree distribution $p_k$ of the network describes the probability
to draw a node with degree $k$ from the network $p_k=Ak^{-\tau}$
where $A$ is a normalization constant such that
$\sum_{i=2}^{k_{max}}p_k=1$.
The main result of~\cite{SBR2004} is the following three-step algorithm:
\emph{Step 1, Content List Implantation:} To insert cached replicas
  of the content, a random walk is performed and each
  peer visited during this random walk receives a copy of the index
  data.  For $\tau=2$, the length of the random walk should be $O(\log N)$.
\emph{Step 2, Query Implantation:} The above process is
  performed for each query.
\emph{Step 3, Bond percolation:} After the query implantation, each node
  that has received a query so far will forward the query to each of its
  neighbors with a probability $q = \gamma q_c$, where $q_c$ is the
  \emph{percolation threshold}: $\langle k\rangle/(\langle k^2\rangle -
  \langle k\rangle)$, and $\gamma$ is a small number greater than unity.

For random power-law networks, there are
two parameters one may control: $\tau$ the exponent of the power-law,
and $k_{max}$ the maximum number of neighbors any node has.  In
this work we will only consider $\tau=2$.
For $\tau=2$, 
$\frac{1}{A} =\sum_{k=1}^{k_{max}} \frac{1}{k^{2}}\approx \pi^2/6\approx 1.6$ \ .
Next we consider the scaling of the average and maximum of bandwidth,
disk and processing requirements for the percolation search.

\subsubsection{Resource requirements}
Given that a node receives a query, with probability $q$, each neighbor sees
that query.  So, the total number of edges to see a query will be $qE$, where
$E$ is the total number of edges.  Since $E = \langle k\rangle N/2$,
and $q\propto q_c = \langle k\rangle/(\langle k^2\rangle -
  \langle k\rangle)\approx \langle k\rangle/\langle k^2\rangle $. we have
that the average bandwidth cost is
$B_{ave} = \frac{RN Eqz}{N} = Rz \frac{\langle k\rangle^2 N}{2\langle k^2\rangle}$.
When $p_{k} = A/k^2$, $\langle k\rangle \approx A\ln k_{max}$, $\langle
k^2\rangle = A k_{max}$, then we have
$B_{ave} \approx RzN \frac{A \ln^2 k_{max}}{2 k_{max}}$.

To compute the maximum bandwidth, we need to look at the highest degree node
and see how many of its neighbors will see the query.  The highest degree
node has $k_{max}$ neighbors, and on average $q k_{max}$ will see the query,
thus: $B_{max} = R z N k_{max} q \approx R z N k_{max} q_c \approx R z N \ln k_{max}$.

For a power-law random network with exponent $\tau=2$, the content is cached on
$\log_2 N$ nodes.  Thus, the average storage requirements are
$D_{ave} = \frac{C z N \log_2 N}{N} = C z \log_2 N = \frac{C z \ln N}{\ln 2}$.

To compute the storage required for the maximum node is more involved.  We
model a random walk on a random network as each step selecting a random node
of degree $m$ with probability $mp_{m}/\langle k\rangle$.  The probability
of selecting a node of the highest degree is $P_s = \frac{k_{max}A}{\langle
k\rangle k_{max}^2}=(k_{max}\ln k_{max})^{-1}$.  We assume that we select
each of the nodes of degree $k_{max}$ with equal probability, so the
probability we select each one of them is: $Q_s = P_s \left(N p(k_{max})\right)^{-1}$.
There are $\log_2 N$ steps, so the number of content caches
that make it to highest degree nodes is $F_{max} = Q_s \log_2 N = 
\frac{k_{max}\ln N}{\ln 2 AN\ln k_{max}}$.
Now we can compute the maximum storage requirements:
\[
D_{max} = CzN F_{max} = \frac{Cz k_{max}\ln N}{A\ln 2\ln k_{max}}
\]

As before, we assume that the search time is linear in the number of items
stored at each node.  Since we have already computed the number of items
stored at each node, we have:
\begin{eqnarray*}
P_{ave} &=& RN D_{ave}(f/z) = RCfN \ln N\frac{1}{\ln 2}\\
P_{max} &=& RN D_{max}(f/z) = RCfN \frac{k_{max} \ln N}{A\ln 2\ln k_{max}}
\end{eqnarray*}

To reduce $P_{max}$ we need to reduce $k_{max}$, but that will increase
$B_{ave}$.

\subsubsection{Trade-offs and numerical values}

In the percolation search, like the super-node system, we can decrease
the average bandwidth required at the expense of increasing the maximum
processor utilization.  Using the percolation search
$B_{ave}P_{max} = fzCR^2N^2 \frac{\ln k_{max} \ln N}{2}$, which is similar
to the super-node architecture except with some logarithmic factors.
In order to minimize average bandwidth, we should choose $k_{max}$ to be as
large as possible.  In general, the percolation search algorithm 
behaves 
like
the ideal super-node algorithm with $s\approx 1/k_{max}$.

To compare to the super-node case, we choose $k_{max} = \sqrt{N}$,
$N=2^{19}$ and the values of $R,C,f,z$ from Table
\ref{tab:nomenclature} and using the same constants we assumed in
\ref{sec:sn_num}, which means our highest degrees are comparable to
super-nodes.  The results for this case are summarized in the following table:

\begin{tabular}{lll}
$B_{ave}=$ & $\frac{A}{8} Rz \sqrt{N}\ln N$ & $\approx 70 B/s = 560 bps$\\
$B_{max}=$ & $\frac{1}{2}Rz N \ln N $ &$\approx 330KB/s = 2.7 Mbps$\\
$D_{ave}=$ & $Cz \frac{\ln N}{\ln 2}$ &$\approx 300 KB$\\
$D_{max}=$ & $Cz\sqrt{N}\frac{2}{A\ln 2}$ &$\approx 52 MB$\\
$P_{ave}=$ & $RCf N\frac{\ln N}{\ln 2}$ &$\approx 7.9M FlOp/s$\\
$P_{max}=$ & $RCf N\sqrt{N}\frac{2}{A\ln 2}$ &$\approx 1.4 G FlOp/s$
\end{tabular}

For $D_{max}$ and hence, $P_{max}$ there is a constant factor overhead
of $2/(A\ln 2)\approx 4.6$ when compared to the ideal super-node case.
For all other metrics, there is an $O(\ln N)$ overhead for using the
percolation search, however, for networks of size $N=2^{19}$, due to
the division by a constant, the difference is not very great.

\subsubsection{Simulation results}
\label{sec:observations}

Our simulations use the Netmodeler package\cite{Netmodeler}.
We insert 1000 content objects at uniformly
selected nodes on a power-law network with $\tau = 2$, and then make 1000 queries from uniformly selected
nodes.  We are particularly concerned with the maximum demands made
on any node.
Some results are in the following table. BW
is the total number of times each query is copied to search
the entire network, and Max $C/O$ is the average of the
maximum CPU where $1$ is the cost to evaluate the query, per
content object in the network.

\begin{tabular}{l|l|l|l|l|l}
Nodes & $q$ & ttl & Hit-rate & BW & Max $C/O$ \\
\hline
$2^{19}$ & 0.01 & 20 & 0.961 & 10,428 & 0.0075\\
$2^{20}$ & 0.01 & 21 & 0.966 & 21,045 & 0.0042
\end{tabular}

We see that for the case
on 1000 content objects, the maximum node had to search $7.5$ and $4.2$
objects for each query on average, for the cases of $2^{19}$ and $2^{20}$
nodes respectively.  To scale these results up to our assumptions
of $20$ content objects per node.
Additionally, the total query byte rate rate will be $z N R =
800 \times 2^{19}\times 1.2 \times 10^{-4}\approx 48,500 Bps = Q$.
For $N=2^{19}$ we have:
\begin{eqnarray*}
B_{ave} &=& \frac{10,428}{2^{19}}Q = 7.7 Kbps\\
D_{max} &=& 13,280\times 2^{19}\times \frac{7.5}{1000} = 52.21 MB\\
P_{max} &=& 20\times 2^{19}Q\frac{332}{800}\frac{7.5}{1000} = 1.6 GFlOp/s 
\end{eqnarray*}
The above parameters are a relatively close match to the predictions
of the previous section. For $N=2^{20}$ we have $Q' = 2Q$:
\begin{eqnarray*}
B_{ave} &=& \frac{21,045}{2^{20}}Q' = 15.6 Kbps\\
D_{max} &=& 13,280\times 2^{20}\times \frac{4.2}{1000} = 58.5 MB\\
P_{max} &=& 20\times 2^{20}Q'\frac{332}{800}\frac{4.2}{1000} = 3.6 GFlOp/s 
\end{eqnarray*}
The other parameters such as $D_{ave}, P_{ave}$ and $B_{max}$ are not dependent
on the nonlinearities of percolation, and as such match predictions
of the previous section.

Our simulations verify that the average bandwidth required is much 
less than analog modems can provide and the maximum processing
requirements are met by one modern desktop CPU.

\section{Comparison of unstructured to structured image search}
\label{sec:compare_utos}

In order to compare unstructured search with current DHT-based
approaches, we took PRISM~\cite{SGE2005} as a base for
comparison. PRISM is a recent system with a clear focus on similarity
queries over high-dimensional vectors.

The PRISM paper also describes load balancing between PRISM peers.
However, within the following, we will consider PRISM without load
balancing, as \emph{perfect} load balancing would amount to all peers
carrying the average load.

\begin{figure}[t]
  \centering
  \includegraphics[width=7cm]{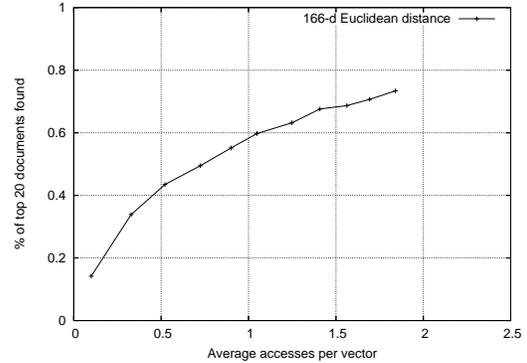}
  \caption{The fraction of 20-NN found plotted against the fraction of the total collection visited in ``Vanilla'' PRISM. }
  
  \label{fig:prism:recall:vs:collection}
\end{figure}

\begin{figure}[t]
  \centering
  \includegraphics[width=7cm]{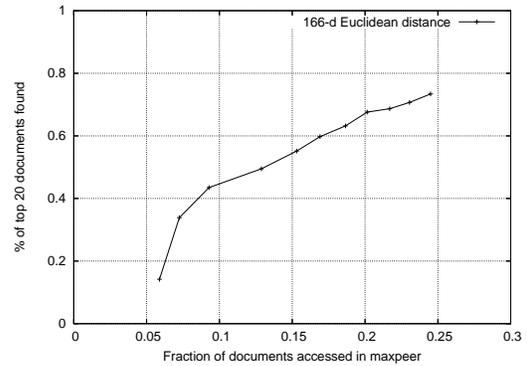}
  \caption{The fraction of 20-NN found plotted against the load on the
    most solicited peer in ``Vanilla'' PRISM without load balancing.
  }
  \label{fig:prism:recall:vs:load:on:maxpeer}
\end{figure}

The performance metrics for the vanilla PRISM algorithm
\emph{without} load balancing. Are given in the table below.
  We chose visiting all 11 reference pairs for our calculation. 
  Again $N=2^{19}$ and the values of $R,C,f,z$ from Table \ref{tab:nomenclature}.
\begin{tabular}{lll}
$B_{ave}=$ & $21\cdot \frac{RNz}N$ &$ =  2.02 B/s = 16.1 bps$\\
$B_{max}=$ & $\frac 14 RNz$ & $=  12,600B/s = 100 kbps$\\
$D_{ave}=$ & $11Cz$ & $ = 176KB$\\
$D_{max}\approx$&  $0.1CzN$ & $\approx 840MB$\\
$P_{ave}\approx$ & $1.8 RCf N $ &$= 670 kFlOp/s$\\
$P_{max}\approx$ & $0.25 RCf N^2$ & $= 55G FlOp/s$
\end{tabular}
\label{tab:prism_perf}

As Figs.  \ref{fig:prism:recall:vs:collection} and
\ref{fig:prism:recall:vs:load:on:maxpeer}, as well as
Tab.~\ref{tab:prism_perf} present our experiments with a simulation of
PRISM.  Without load balancing, PRISM behaves to quite an extent like
a client/server system: most load hits few servers. Little load is
distributed, so there is low communication cost.  For Euclidean
distance, PRISM presents ---if any--- only small advantages over the
super-peer network. In our current setup, in order to find 75\% of the
top 20 documents, we have to visit each vector more than one time on
average, \ie\ PRISM performs worse than random search. If one wants to
push the recall to $100\%$ (as good as the super peer method), each
item is considered even more times on average, incurring a clear
efficiency penalty with respect to a full scan.

The second finding is that the distribution of data items over peers is heavily
skewed, emphasizing the need for load balancing as proposed in~\cite{SGE2005}.
Our experiment used $32\times 32 = 1024$ pairs.
In these experiments, the first 5 most used pairs account for more than 10\% of the
traffic, the first 15 pairs account for more than 25\% of the traffic, and the
first 60 pairs account for more than 50\% of the traffic.  To highlight this
fact, Tab.~\ref{tab:prism_perf} shows PRISM without load balancing. Here, PRISM
functions almost in a client/server-alike fashion.  Please note that with the
proper use of load balancing, PRISM would thus much behave like a super-peer
network discussed in Section \ref{sec:supernode}, but with slightly higher
load for the super-peers.

The 
third
finding, finally, should spawn a series of new experiments:
The performance of systems like PRISM depends also on data set and
distance measure. However, most of the distributed indexing literature
is fixated on the Euclidean metric and similar distance measures. Our
experiments show that it is clearly worthwhile to investigate deeper
into the performance of such systems when using more diverse distance
measures.



\section{Conclusion}
Summarizing, the performance of structured and unstructured systems
seem to be pretty close in our application domain, while unstructured
systems have the advantage of being more flexible with respect to the
queries they allow.  We should mention
that this conclusion is similar to the recent paper of \cite{YDR2006}.

{
\small
\bibliographystyle{latex8}
\bibliography{bib}

\begin{thebibliography}{10}\setlength{\itemsep}{-1ex}\small

\bibitem{Netmodeler}
P.~O. Boykin.
\newblock Netmodeler complex network simulator.
\newblock http://boykin.acis.ufl.edu/wiki/index.php/Netmodeler.

\bibitem{GGP2002}
A.~Graham, H.~Garcia-Molina, A.~Paepcke, and T.~Winograd.
\newblock Time as essence for photo browsing through personal digital
  libraries.
\newblock In {\em JCDL '02: Proceedings of the 2nd ACM/IEEE-CS joint conference
  on Digital libraries}, pages 326--335, New York, NY, USA, 2002. ACM Press.

\bibitem{KZ2004}
M.~Kleis and X.~Zhou.
\newblock A placement scheme for peer-to-peer networks based on principles from
  geometry.
\newblock In {\em Proceedings of the Fourth International Conference on
  Peer-to-Peer Computing}, pages 134--141. IEEE Press, 2004.

\bibitem{KOR2000}
E.~Kushilevitz, R.~Ostrovsky, and Y.~Rabani.
\newblock Efficient search for approximate nearest neighbor in high dimensional
  spaces.

\bibitem{MaM2002}
P.~Maymounkov and D.~Mazieres.
\newblock {Kademlia: A peer-to-peer information system based on the XOR
  metric}.
\newblock In {\em Proc of IPTPS}, 2002.

\bibitem{NWS2003}
W.~Nejdl, M.~Wolpers, W.~Siberski, C.~Schmitz, M.~Schlosser, I.~Brunkhorst, and
  A.~L\"oser.
\newblock Super-peer-based routing and clustering strategies for rdf-based
  peer-to-peer networks.
\newblock In {\em Proc. of the Intl. World Wide Web Conf.}, 2003.

\bibitem{NgS2003}
C.~H. Ng and K.~C. Sia.
\newblock Bridging the {P2P} and {WWW} divide with {DISCOVIR} - distributed
  content-based visual information retrieva.
\newblock In {\em Poster Proc. of The 11th Interational World Wide Web Conf.
  \em{to be published}}, 2003.

\bibitem{SGE2005}
O.~D. Sahin, A.~Gulbeden, F.~Emekci, D.~Agrawal, and A.~E. Abbadi.
\newblock {PRISM}: indexing multi-dimensional data in {P2P} networks using
  reference vectors.
\newblock In {\em MULTIMEDIA '05: Proceedings of the 13th annual {ACM}
  international conference on Multimedia}, pages 946--955, New York, NY, USA,
  2005. ACM Press.

\bibitem{SBR2004}
N.~Sarshar, P.~O. Boykin, and V.~P. Roychowdhury.
\newblock Percolation search in power law networks: making unstructured
  peer-to-peer networks scalable.
\newblock In {\em Proceedings of Fourth International Conference on
  Peer-to-Peer Computing}, pages 2--9. IEEE, August 2004.

\bibitem{SWS2000}
A.~W. Smeulders, M.~Worring, S.~Santini, A.~Gupta, and R.~Jain.
\newblock Content based retrieval at the end of the early years.
\newblock {\em T-PAMI}, 22(12):1349--1380, 2000.

\bibitem{Smi1997}
J.~Smith.
\newblock {\em Integrated Spacial and Feature Image Systems: Retrieval,
  Compression and Analysis}.
\newblock PhD thesis, Graduate School of Arts and Sciences, Columbia
  University, 2960 Broadway, New York, NY, USA, 1997.

\bibitem{SMN2001}
I.~Stoica, R.~Morris, D.~Karger, F.~Kaashoek, and H.~Balakrishnan.
\newblock Chord: {A} scalable {Peer-To-Peer} lookup service for internet
  applications.
\newblock In {\em Proc. ACM SIGCOMM Conf.}, San Diego, CA, USA, 2001.

\bibitem{TXM2002}
C.~Tang, Z.~Xu, and M.~Mahalingam.
\newblock {pSearch}: Information retrieval in structured overlays.
\newblock In {\em First Workshop on Hot Topics in Networks (HotNets-I)},
  Princeton, NJ, 2002.

\bibitem{WSB1998}
R.~Weber, H.-J. Schek, and S.~Blott.
\newblock A quantitative analysis and performance study for similarity-search
  methods in high-dimensional spaces.
\newblock In {\em Proc. Intl. Conf. on VLDB}, New York, USA, 1998.

\bibitem{YaG2003}
B.~Yang and H.~Garcia-Molina.
\newblock Designing a super-peer network.
\newblock In {\em IEEE International Conference on Data Engineering, 2003},
  2003.

\bibitem{YDR2006}
Y.~Yang, R.~Dunlap, M.~Rexroad, and B.~F. Cooper.
\newblock Performance of full text search in structured and unstructured
  peer-to-peer systems.
\newblock In {\em Proceedings of the 25th Conference on Computer Communications
  (INFOCOM 2006)}. IEEE Press, 2006.
\newblock to appear.

\end{thebibliography}
}
\end{document}